# Improving Emergency Department ESI Acuity Assignment Using Machine Learning and Clinical Natural Language Processing

Oleksandr Ivanov, PhD – Mednition Inc., Ukraine

Lisa Wolf, PhD, RN, CEN, FAEN – Emergency Nurses Association, USA

Deena Brecher, MSN, RN, ACNS-BC, CEN, CPEN, FAEN – Mednition Inc., USA

Erica Lewis, MD – El Camino Hospital, USA

Kevin Masek, MD – San Mateo Medical Center, USA

Kyla Montgomery – Mednition Inc., USA

Yurii Andrieiev – Mednition Inc., Ukraine

Moss McLaughlin – Mednition Inc., USA

Stephen Liu, MD, FACEP – Adventist Health White Memorial, USA

Robert Dunne, MD, FACEP – Ascension Health, USA

Kevin Klauer, DO, EJD, FACEP, FACOEP – American Osteopathic Association, USA

Christian Reilly – Mednition Inc., USA

Corresponding author: Christian Reilly, creilly@mednition.com

Author Credit

**Oleksander Ivanov**: conceptualization, methodology, data curation, formal analysis, Investigation, Writing – Original Draft, Review and Editing; **Lisa Wolf**: Writing- Original Draft, Review and Editing; **Deena Brecher**: Writing- Review and editing; **Kevin Masek**: Writing- Review and editing; **Erica Lewis**: Writing- Review and editing; **Kyla Montgomery**: Formal analysis, review and editing; **Yuri Andrieiev**: Formal analysis, review and editing; **Moss McLaughin:** Formal analysis, review and editing; **Stephen Liu**: senior clinical review; **Robert Dunne**: senior clinical review; **Kevin Klauer**: senior clinical review;

**Christian Reilly**: Conceptualization, Methodology, Formal Analysis, Investigation, Review and editing, Supervision, Project Administration

# Abstract

**Background**

Effective triage is critical to mitigating the effect of increased volume by accurately determining patient acuity, need for resources, and establishing effective acuity-based patient prioritization. The purpose of this retrospective study was to determine whether historical EHR data can be extracted and synthesized with clinical natural language processing (C-NLP) and the latest ML algorithms (KATE) to produce highly accurate ESI predictive models.

**Method**

An ML model (KATE) for the triage process was developed using 166,175 patient encounters from two participating hospitals. The model was then tested against a gold set that was derived from a random sample of triage encounters at the study sites and correct acuity assignments were recorded by study clinicians using the Emergency Severity Index (ESI) standard as a guide.

**Result**

At the two study sites, KATE predicted accurate ESI acuity assignments 75.9% of the time, compared to nurses (59.8%) and average individual study clinicians (75.3%). KATE accuracy was 26.9% higher than the average nurse accuracy (p-value < 0.0001). On the boundary between ESI 2 and ESI 3 acuity assignments, which relates to the risk of decompensation, KATE was 93.2% higher with 80% accuracy, compared to triage nurses with 41.4% accuracy (p-value < 0.0001).

**Conclusion**

KATE provides a triage acuity assignment substantially more accurate than the triage nurses in this study sample. KATE operates independently of contextual factors, unaffected by the external pressures that can cause under triage and may mitigate the racial and social biases that can negatively affect the accuracy of triage assignment. Future research should focus on the impact of KATE providing feedback to triage nurses in real time, KATE's impact on mortality and morbidity, ED throughput, resource optimization, and nursing outcomes.

## 1. Background

Emergency department (ED) use has increased by 35% in the past 20 years, while the number of operating EDs has gone down by 11%, affecting approximately 145M visits in 2016 [1, 2]. Effective triage is critical to mitigating the effect of these problems by accurately determining patient acuity, need for resources, and setting the patient on an optimal clinical workflow path to deliver the safest and highest level of care at the lowest possible cost, primarily based on acuity.

In the US, the most widely used triage tool is the Emergency Severity Index (ESI); a 5-point scale, with 1 being "emergent" and 5 being "nonurgent", assigned by a triage nurse and based on risk of decompensation and anticipated resource utilization [3]. In a report in 2012, 70% of large hospitals and teaching hospitals reported using ESI, while a three-tiered system was still common in community and small hospitals [4].

While the ESI classification system may have clear guidelines for acuity assignments using patient vital signs, its classification methodology, which also factors in expected resource needs implicitly leaves a lot of room for clinician interpretation. Accuracy therefore relies heavily on individual triage clinician judgment and experience. Cognitive biases that can impede clinical decision-making are well documented in the physician literature [5, 6] and the nursing literature [7, 8] including overconfidence, premature closure, the anchoring effect, information and availability bias, and tolerance to risk. Other challenges to accurate acuity assignment can occur when environmental conditions such as overcrowding and boarding, make it difficult to bring patients back to a treatment space [9]. Fry and Burr [10] and Chung [11] note that accuracy was affected by interruptions in care delivery, lack of knowledge, and time constraints; it was also reported that triage clinicians manipulated the triage system to speed up or delay care. In other words, triage clinicians respond not just to the patient before them, but to the environment around them to the potential detriment of the patient [9]. Race and gender bias, specifically, can play a role in incorrect acuity assignment, even with dangerous presentations and abnormal vital signs [12-18], setting patients on a suboptimal trajectory of care [19].

Researchers report nurse triage accuracy of 59.2% in a classroom-like setting [20], and 59.6% in similar conditions, despite a good inter-rater reliability and high nurse confidence in ability to apply the ESI algorithm [21]. Challenges in assigning ESI 1 or 2 (resuscitative, emergent) and 5 (non-urgent) are common, with specific difficulties reported in differentiating between ESI 2 (unstable) and 3 (stable) [22].

Given these challenges, machine learning (ML) approaches have been proposed to aid clinicians in various patient risk assessments. ML has been used to predict in-hospital mortality, critical care (admission to an intensive care unit and/or in-hospital death) and hospitalization (direct hospital admission or transfer) in adults [23-26] and children [27]. ML has also been used to predict hospitalization outcomes [28-30], and predict composite risk outcome [31], defined as an occurrence of mortality, admission to the ICU, or direct transport to the or cardiovascular catheterization suite. Rajkomar used ML to predict in-hospital mortality, readmission, prolonged length of stay and discharge diagnoses [24]. Other applications of statistical forecasting in EDs are reviewed by Gul and Celic [32].

Algorithms based on the results of ML predictions of hospitalization, critical care, and mortality have also been developed to predict triage acuity. Dugas et al. applied thresholds to predicted probabilities of composite outcome to predict triage acuity [31]. Similarly, Levin et al. developed an E-triage algorithm with three ML models to predict critical care outcomes, emergent procedures, and hospitalizations, then applied thresholds on these predictions to determine triage acuity [33].

To the best of our knowledge, no machine learning model has used triage acuity score (ESI) as labels to predict ESI and measure model performance. Therefore, it is not known if an ML model can accurately predict ESI triage acuity using only the information available to the triage nurse at the time of triage.

The purpose of this retrospective study was to determine whether historical EHR data can be extracted and synthesized with clinical natural language processing (C-NLP) and the latest ML algorithms (KATE) to produce highly accurate ESI predictive models.

## 2. Method

**2.1. Study Sites**

Site A is a community emergency department in an urban setting in the Western United States, with 65,000 annual visits. Site B is a level 1 trauma center in an urban setting in the Midwestern United States, with approximately 110,000 ED annual visits. Nurses in both sites performing triage had received training on ESI and triage processes as part of their employment. The study dataset included 88,237 triage records from Site A (visits between February 2015 to November 2016) and 77,938 from Site B (visits between October 2015 to October 2016). Demographics for the triage records are presented in Table 1.

| Table 1. Study site demographic and ESI acuity distributions | | |
|---|---|---|
| **Demographics/Distribution** | **Site A** | **Site B** |
| Total number of records | 88,237 | 77,938 |
| Number of males | 38,159 | 33,482 |
| Number of females | 49,420 | 44,455 |
| Number of pediatric records (age below 18) | 26,772 | 14,027 |
| Number of adult records (age 18 and above) | 60,815 | 63,911 |
| ESI 1 | 191 (0.22%) | 502 (0.64%) |
| ESI 2 | 8,486 (9.62%) | 10,877 (13.96%) |
| ESI 3 | 37,730 (42.76%) | 44,565 (57.18%) |
| ESI 4 | 35,531 (40.27%) | 19,065 (24.46%) |
| ESI 5 | 5,715 (6.48%) | 2,929 (3.76%) |
| Missing ESI | 584 (0.66%) | - |
| Formal Education in Triage | Yes | Yes |
| ESI 1-5 distribution: Nurse assigned ESI acuity in triage | | |

## 2.2. Data Collection and Conditioning

Human subjects protection and conflicts of interest: Prior to data collection, IRB exemption was obtained for both sites from the Western Institutional Review Board (WIRB, OHRP/FDA Parent Organization number: IORG0000432, IRB registration number: IRB00000533). This work was conducted and independently funded by Mednition Inc. a private corporation. All Authors that are with the Mednition organization are employees; non Mednition authors are contract employees of Mednition with the exception of Robert Dunne, MD.

All protected health information (PHI) was redacted from the datasets, in accordance with 45 CFR § 164.514 for de-identifying PHI to safe-harbor standards. De-identified raw text files were mapped and consolidated to a multi-hospital hierarchical data model, preserving the differences between sites. From the initial 166,175 triage records 19,123 records were removed for patients under the age of 1 year, and/or if the record had missing ESI acuity, reason for visit, or four or more vital signs. This data filtering produced the final dataset of 147,052 usable encounters.

**2.3. Machine Learning**

2.3.1. Clinical Natural Language Processing (C-NLP)

Extraction of clinical terms from patient record free text is a prerequisite to form a complete understanding of each patient and can enhance ML-based clinical decision models [28]. This has been a primary challenge in building ML-based clinical decision support tools for clinicians which leverage clinical raw text evidence. The challenge is to accurately understand the individual's information as documented then aggregate that understanding across the research dataset. We developed C-NLP technology to accurately extract medical terms from free text.

In C-NLP we use the following steps to process raw text:

1. Sentences tokenization
2. Words tokenization
3. Text normalization
4. Part of speech tagging
5. Chunking
6. Extraction of clinical terms

Steps 1-5 are done with OpenNLP Java library. Extraction of clinical terms in step 6 is done with following steps:

1. Noun phrases (NP) are extracted from chunker (step 5)
2. Text in each NP is permuted in all combinations for all phrases
3. All text combinations are matched against UMLS dictionary [34], clinical terms are extracted based on

matching to UMLS terms.

4. For each medical term unique UMLS code (CUI) is extracted and used as features.

To evaluate C-NLP technology, we randomly sampled 800 medical records. For each medical record two independent trained reviewers evaluated the relevance and accuracy of each tag predicted by C-NLP. The performance of C-NLP is presented in Table 2. A more detailed analysis of C-NLP performance is presented in Supplementary Table 1.

**Table 2.** Performance and 95% confidence intervals (in parentheses) of C-NLP for 800 randomly sampled medical records

| Score | Value |
| --- | --- |
| Number of clinical terms | 9506 |
| Accuracy (95% CI) | 0.9847 (0.9822 - 0.9871) |
| F1 score (95% CI) | 0.9923 (0.991 - 0.9935) |
| Sensitivity (95% CI) | 0.997 (0.9957 - 0.9981) |
| Precision (95% CI) | 0.9877 (0.9853 - 0.9897) |

Features extracted for this study from free text using C-NLP are summarized in Table 3.

2.3.2. Feature Engineering

In this study, numerical, categorical, and free text data was used. Numerical data is represented by age, vital signs, pain scores, Glasgow coma scale score, and other triage assessments. Numerical data was transformed into features after removing outliers. Categorical data is represented by sex, arrival mode, arrived from, family history, social history, and risk factors such as alcohol and drug abuse. Overall, information that was available at triage was used as a data source. Post-triage data, such as labs or vitals signs after triage, was not used as the data source. Clinical terms were extracted from chief complaints, and patient histories (medical, social, surgical, and medication data) using C-NLP.

Clinical feature engineering was undertaken to derive new composite features from the existing EHR data and public datasets, which improved predictive value for ESI triage acuity assignment. The following feature engineering algorithms were applied:

- UMLS dictionaries of clinical terms were used to derive consolidated features based on features extracted from reason for visit using C-NLP. UMLS is a collection of dictionaries, many of which have a primary term for medical terms [34]. For example, "radiating chest pain" is related to "chest pain". For each clinical term, extracted from reason for visit using C-NLP, we used relationships to derive new features [34].
- ESI 1 and ESI 2 features were created based on the presence of high-risk presentations referenced in the ESI Handbook [3].
- Social and environmental risk factors were binned into risk and non-risk categories.
- Duration of symptoms features were created based on time references in reason for visit (e.g., hours, days, weeks).
- Count of number of features extracted from reason of visit using C-NLP.
- Pain above acceptable level for each patient.
- Count of Number of NAs in vital signs.
- Count of number of vital signs in the risk zone.

Features with low frequency were removed from the feature set.

A summary of the feature engineering to improve modeling and subsequent clinical accuracy of KATE triage prediction is presented in Table 3. A model feature is equivalent to a clinical data point. For reference, "chest pain", "denies chest pain", and "history of myocardial infarction" would represent three unique features in KATE.

The concept of time was extracted using regexes and binned to categories: seconds, minutes, hours, days, weeks, months and years. For example, "a few days", "3 days", or "three days" are all binned into the "days" category. Duration categories were used as features.

**Table 3.** Statistics of features used in training and validation of machine learning models

| Clinical Natural Language Processing (C-NLP) features overview | Count |
|---|---|
| Total patient encounters used for model training and validation | 147,052 |
| Total free text words processed | 12,158,342 |
| Total clinical features extracted from free text | 1,880,841 |
| Average free text clinical features per triage encounter | 12.79 |

**KATE features overview**

| Category | Description | Number of unique features | Total count of feature values |
|---|---|---|---|
| Total extracted features | All EHR Study data including C-NLP extracted free-text features as input to feature engineering | 45,928 | ~ 10.1 Billion |
| Total features after feature engineering | Processing, consolidation, and enhancement of raw data, input to KATE | 26,332 | ~ 9.9 Billion |
| Total features selected by model | Features selected by model with positive predictive power (e.g. non-zero gain) | 7,679 | ~ 9.1 Billion |
| KATE engineered features | Subset of model features with positive predictive power added through feature engineering | 3,554 | ~ 3.5 Billion |

**Study feature predictive power (gain) overview**

| Item | Description | Total gain | Average gain per Feature |
|---|---|---|---|
| All features | Total model selected features | 100% | 0.0013% |
| Study EHR features | Model features found in raw EHR data | 53.7% | 0.0012% |
| KATE engineered features | Study team feature engineered features | 46.3% | 0.0014% |

2.3.3. Machine Learning Algorithms

XGboost was selected as the ML algorithm used for this study. XGboost is a method from the gradient tree boosting family. Gradient boosting is a method of sequential building of decision trees in which each subsequent tree is built on the subset of data where previous trees made the most of mistakes in classification [35]. XGboost was designed to have an efficient model training performance for large sparse datasets [35].

2.3.4. Software

Java 8.0 and OpenNLP Java library were used to develop C-NLP. Python 2.7 was used for pipeline development and machine learning. XGboost 0.8 library was used to build KATE. Sklearn and Scipy libraries were used for model evaluation and statistical analysis.

**2.4. Clinically Verified Training Records**

Given the high nurse triage error rate reported in literature [20, 21] and found in this study (Section 3.1), triage records were reviewed by the study clinical team, three practicing experts in emergency medicine, in order to correct potential nurse errors in ESI assignment. The expert clinicians were a doctorally prepared emergency nurse with nationally recognized expertise in emergency department triage, and two emergency physicians with an average of 10 years' experience in emergency medicine. We used both physicians and nurses in this process for interdisciplinary expertise and also because accuracy in ESI triage is not substantially different between the two groups [36]. The physicians received education about the clinical decision-making process of nurses performing triage and training on the ESI algorithm. Disagreements about ESI acuity levels were resolved using the ESI handbook. A total of 19,652 records were reviewed and relabeled by one or more clinicians. Triage records for clinician review were chosen on the basis of disagreement of KATE ESI predictions with nurse ESI from the results of five-fold cross-validation. Triage records reviewed by study clinicians were considered verified, the remaining records were considered unverified. For 7,970 (40.54%) verified records study clinicians agreed with nurse-assigned ESI acuity and for 11,682 clinicians changed ESI acuity (59.44%). A confusion matrix for nurse ESI against verified ESI is presented in Supplementary Table 2.

Both verified and unverified records were used as a training set for the KATE model. Gold records were used as a test set since gold records have the highest certainty in ESI labels for evaluation of nurses, the KATE model, and expert clinicians. The results of KATE performance on the gold set with and without verified ESI acuity labels are presented in Supplementary Table 3. The use of verified ESI acuity labels significantly improves KATE performance on the gold set. Specifically, accuracy on all gold set records is improved by 6.75%, and ESI 2 accuracy improved by 17.13%.

**2.5. Results Validation**

Given the high nurse error rate in triage reported in literature [20, 21] and found in this study, to validate the model performance a test set (gold set) was created using random sampling from the overall dataset of 147,052. A random sample of 800 records was drawn with 3.62% margin of error at a 95% confidence level. Hospital assigned acuities were redacted from the gold set during this review. Each gold set record was reviewed independently by three clinicians trained in the ESI methodology. ESI acuity assignments were made prospectively using triage information only. If all three clinicians agreed in their acuity assignment, such acuity was determined to be the correct acuity for this record. For records with disagreements, each case was discussed in a team clinical review, referenced against the ESI Handbook, and a final correct acuity was recorded, or the record was removed from the study gold set (deleted). Of the initial random sample of 800 records, 71 triage records (8.9%) were deleted due to insufficient triage documentation (missing vital signs, lack of a reason for visit or basic patient assessment) or the study clinical team could not reach consensus (Supplementary Table 4). Triage acuity labels in the remaining 729 triage records from the gold set were isolated from the study's machine learning training, and only used for model testing. Acuity assignment accuracy of KATE, site nurses, and study clinicians were tested on the gold set.

## 3. Results

**3.1 Overall Results**

Results of KATE, nurses, and individual study clinicians are presented for two study sites in Tables 4 and 5. KATE demonstrates significantly higher accuracy than nurses both for all gold records from the two study sites and for each triage acuity level. Specifically, for the two study sites KATE accuracy is 75.9%, which is 26.9% higher than average nurse accuracy 59.8% (p-value < 0.0001). For study Site A, KATE accuracy is 23.5% higher than the average nurse accuracy (p-value < 0.0001), and for study Site B, KATE accuracy is 30.8% higher than the average nurse accuracy (p-value < 0.0001). KATE accuracy for the two study sites combined is not significantly different to each of three study clinicians (corresponding p-values are 0.399, 0.632 and 0.055). Importantly, KATE also demonstrates 80% accuracy for the 2/3 triage acuity boundary, whereas ED triage nurses demonstrate 41.4% accuracy (93.2% improvement over nurses). This significant improvement in performance did not come at the cost of a high false positive rate on the 2/3 boundary, as KATE's ESI 3 over triage was 6.9% vs. nurse 7.9% (Table 4). Distributions of triage acuity assignment by nurses, KATE and three study clinicians for triage records from the gold set are presented in Supplementary Table 5.

KATE demonstrates significantly higher AUC (0.849 vs 0.749 correspondingly, Supplementary Table 6), F1

scores (0.738 vs 0.428 correspondingly, Supplementary Table 7), sensitivity (0.695 vs 0.417, Supplementary Table 8), and precision (0.809 vs 0.488, Supplementary Table 9) than nurses both for all gold records from the two study sites.

Table 4. Accuracy and 95% confidence intervals (in parentheses) of KATE, nurses, and three study clinicians for the gold set and two study sites individually.

| Group | Number of triage records | KATE accuracy (95% CI) | Nurse triage accuracy (95% CI) | Clinician 1 accuracy (95% CI) | Clinician 2 accuracy (95% CI) | Clinician 3 accuracy (95% CI) |
|---|---|---|---|---|---|---|
| All records | 729 | 0.759 (0.724 - 0.787) | 0.598 (0.56 - 0.632) | 0.776 (0.744 - 0.804) | 0.768 (0.738 - 0.794) | 0.715 (0.683 - 0.746) |
| Site A | 368 | 0.799 (0.758 - 0.834) | 0.647 (0.595 - 0.696) | 0.745 (0.696 - 0.785) | 0.726 (0.677 - 0.769) | 0.719 (0.667 - 0.762) |
| Site B | 361 | 0.717 (0.665 - 0.759) | 0.548 (0.493 - 0.593) | 0.809 (0.77 - 0.848) | 0.812 (0.767 - 0.85) | 0.711 (0.661 - 0.753) |
| All ESI 1 | 5 | 0.6 (0.2 - 1.0) | 0.0 (0.0 - 0.0) | 0.6 (0.2 - 1.0) | 0.6 (0.2 - 1.0) | 0.4 (0.0 - 0.8) |
| All ESI 2 | 145 | 0.8 (0.724 - 0.869) | 0.414 (0.331 - 0.49) | 0.766 0.697 - 0.835) | 0.779 (0.71 - 0.842) | 0.757 (0.681 - 0.819) |
| All ESI 3 | 277 | 0.827 (0.776 - 0.87) | 0.769 (0.715 - 0.82) | 0.83 (0.783 - 0.87) | 0.845 (0.798 - 0.884) | 0.8 (0.749 - 0.844) |
| All ESI 4 | 210 | 0.757 (0.695 - 0.814) | 0.676 (0.614 - 0.733) | 0.81 (0.757 - 0.862) | 0.795 (0.743 - 0.848) | 0.676 (0.605 - 0.733) |
| All ESI 5 | 92 | 0.5 (0.391 - 0.598) | 0.228 (0.141 - 0.315) | 0.565 (0.457 - 0.663) | 0.467 (0.358 - 0.565) | 0.494 (0.393 - 0.596) |

Group description:
All records - validation records in the gold set from Site A and Site B
All ESI 1-5 - ESI triage acuity from Site A and Site B

KATE demonstrates both significantly lower under triage (9.7%) and over triage (14.4%) compared to nurses (19.8% and 20.4% respectively) from both study sites (Table 4). Hence, nurses demonstrate 104% higher rate of under triage and 41% higher over triage than KATE for two study sites. On the other hand, KATE is not significantly different from the study clinical team with respect to under triage and over triage (11.8% and 12.9% respectively). Nurses demonstrate a very high level of under triage for acuity level 2 (57.9%), whereas KATE and study clinicians demonstrate 20% and 23.3% under triage rate respectively. KATE and the study clinical team also demonstrate lower rate of over triage for acuity level 5 (50% and 49.1% respectively), compared to nurses (77.2%).

KATE, nurses and study clinicians were evaluated for adult (18 years old and over) and pediatric (1 to 17 years old) patients separately. For adult patients, KATE accuracy is 78.5%, nurse accuracy is 61.6%, and three study clinicians 81%, 78.3% and 74.8%, respectively (Supplementary Table 10). For pediatric patients, KATE accuracy is 67.1%, nurse accuracy is 53.9%, and study clinicians 66.5%, 71.9% and 60.4%, respectively (Supplementary Table 11).

**Table 5.** Under triage and over triage rates with 95% confidence intervals (in parentheses) of KATE, nurses, and three study clinicians for the gold set and two study sites individually.

| Group | Number of triage records | KATE under triage (95% CI) | Nurse under triage (95% CI) | Average of clinicians under triage | KATE over triage (95% CI) | Nurse over triage (95% CI) | Average study clinicians over triage |
|---|---|---|---|---|---|---|---|
| All records | 729 | 0.097 (0.077 - 0.119) | 0.198 (0.169 - 0.228) | 0.118 | 0.144 (0.117 - 0.17) | 0.204 (0.174 - 0.233) | 0.129 |
| Site A | 368 | 0.098 (0.065 - 0.128) | 0.193 (0.149 - 0.234) | 0.093 | 0.103 (0.073 - 0.133) | 0.16 (0.12 - 0.198) | 0.178 |
| Site B | 361 | 0.097 (0.067 - 0.125) | 0.202 (0.161 - 0.244) | 0.144 | 0.186 (0.147 - 0.224) | 0.249 (0.202 - 0.296) | 0.079 |
| All ESI 1 | 5 | 0.4 (0.0 - 0.8) | 1.0 (1.0 - 1.0) | 0.467 | - | - | - |
| All ESI 2 | 145 | 0.2 (0.131 - 0.262) | 0.579 (0.49 - 0.655) | 0.233 | 0.0 (0.0 - 0.0) | 0.007 (0.0 - 0.021) | 0.000 |
| All ESI 3 | 277 | 0.105 (0.065 - 0.141) | 0.152 (0.105 - 0.195) | 0.124 | 0.069 (0.04 - 0.098) | 0.079 (0.047 - 0.112) | 0.051 |
| All ESI 4 | 210 | 0.052 (0.024 - 0.086) | 0.062 (0.029 - 0.095) | 0.073 | 0.19 (0.138 - 0.243) | 0.262 (0.205 - 0.319) | 0.167 |
| All ESI 5 | 92 | - | - | - | 0.5 (0.391 - 0.598) | 0.772 (0.674 - 0.848) | 0.491 |

Group description:
All records - validation records in the gold set from Site A and Site B
All ESI 1-5 - ESI triage acuity from Site A and Site B

**3.2 High-Risk Presentation Results**

KATE and nurse performance were also evaluated on high-risk presentations, which may be associated with ESI 1 or 2 and referenced in the ESI Handbook [3]. For high-risk presentations that appeared in the gold set with frequency 5 or greater, the results are presented in Table 6. KATE demonstrates significantly higher accuracy compared to nurses for these selected high-risk presentations.

Table 6. Accuracy and 95% confidence intervals (in parentheses) of KATE and nurses for a selection of high-risk presentations.

| ESI Handbook - ESI 2 criteria | Count in gold set ($n=729$) | KATE gold accuracy (95% CI) | Nurse gold accuracy (95% CI) |
|---|---|---|---|
| Significant tachycardia | 35 | 0.971 (0.914 - 1.0) | 0.371 (0.2 - 0.542) |
| Altered level of consciousness | 25 | 0.84 (0.68 - 0.96) | 0.36 (0.16 - 0.56) |
| Systemic inflammatory response syndrome | 24 | 0.75 (0.583 - 0.917) | 0.417 (0.208 - 0.625) |
| Hypotension | 10 | 1 (1.0 - 1.0) | 0.2 (0.0 - 0.5) |
| Symptomatic hypertension | 10 | 0.9 (0.7 - 1.0) | 0.8 (0.5 - 1.0) |
| Recent seizure | 8 | 0.875 (0.625 - 1.0) | 0.125 (0.0 - 0.375) |
| Active chest pain | 6 | 0.833 (0.5 - 1.0) | 0.5 (0.167 - 0.833) |
| Suicidal ideation | 6 | 1 (1.0 - 1.0) | 0.833 (0.5 - 1.0) |

## 4. Discussion

The purpose of this retrospective study was to determine whether historical EHR data could be extracted and synthesized with clinical natural language processing (C-NLP) and the latest ML algorithms (KATE) to produce highly accurate ESI predictive models. Accuracy in the initial assignment of emergency department patient acuity is a critical function, with implications for clinical trajectory and resource deployment. Under triage has been associated with a significantly higher rate of admission and critical outcome, while over triage is associated with a lower rate of both [37]. Although over triage can result in overutilization of resources extending lengths of stay, the pervasiveness of under triage, seen in both our sample and in other studies, has potentially serious consequences in patient situations due to delays in care [19].

This study has demonstrated with retrospective data that highly accurate predictions are feasible. We found that KATE predicted accurate ESI acuities using only the information available to the triage clinician at the time of triage (e.g. no final disposition or diagnosis was used) 75.9% of the time, as compared with nurses (59.8%) and study clinicians (75.3%). KATE outperformed nurses on all acuity levels for both study sites. KATE also demonstrates superior performance in accuracy compared to nurses for pediatric, 67.1% and 53.9% respectively, and adult patients, 78.5% and 61.6% respectively. Further research is required to improve KATE performance for pediatric patients.

Importantly, KATE accuracy on the border of ESI 2 and 3 (greater risk of decompensation) was 80% across both sites vs. 41.38% for ED nurses. Besides high accuracy overall for each acuity level, KATE also demonstrates high accuracy for common high-risk presentations used as examples in the ESI Handbook. (Table 6). Not only do ED nurses in this study demonstrate a 40% error rate on average, the errors are clustered for specific presentations, including those that would be high-risk.

There are both individual and environmental factors that can influence the perception of acuity in triage. Individual factors including knowledge deficits and implicit bias can impede accuracy in triage acuity assignment [14-16, 18, 37]; in the design and training of KATE, ethnicity, socio-economic status, and geographic data were not used as model features, potentially mitigating these biases. We specifically excluded race and socio-economic status as factors both because reliance on these as predictors have been shown to decrease accuracy [14-18], and there is no biological or clinical basis on which to weight these factors. [38]

External or environmental factors such as crowding, chaos, cognitive bias, and time pressure can also affect the ability of the triage nurse to accurately perceive acuity [3, 9, 39, 40]. KATE evaluates each patient individually, and thus emergency department conditions do not affect the perception of acuity; in a chaotic ED, the clinical support provided by KATE may mitigate errors due to interruptions of the triage process.

Additionally, there are potential real-time benefits to a clinical decision support aid; researchers report that under triage by nurses fell from 26.3% to 9.3% after an ESI refresher course was provided [41]. Because KATE could provide real-time feedback on clinical decision-making at triage, the program itself may also be useful as a self-directed educational process, improving nursing accuracy.

To the best of our knowledge, KATE is the first machine learning model that was trained on patient triage records and uses acuity score (ESI) as labels to predict ESI. Other approaches that use machine learning to predict triage acuity, such as Dugas et al. [31] and Levin et al. [33] focus on an ML model to predict mortality, admission or critical care outcome and then apply ad-hoc thresholds to determine ESI. The limitation of these approaches is that although predicted outcomes are correlated with high acuity, there are specific clinical presentations such as hypoglycemic event, anaphylaxis, or opioid overdose for which this may not be the case, nor do the algorithms evaluate resources explicitly or implicitly, which is important for distinction between ESI 3, 4 and 5. In addition, cutoff thresholds are not known beforehand, and their optimization is not as efficient and hospital-independent as the ML approach, which was used in KATE.

Valuable information about patient presentation, e.g. reason for visit, comes in the form of free text. Extracting relevant clinical information from reason for visit is crucial for any ML predictive model in ED. Several approaches have been applied to extract information from reason for visit. Zhang et al. extracted significant words for disposition using chi-square test [28]. Rajkomar et al. used individual words [24]. Hong et al. extracted 200 most frequent reasons for visit and used them as categories [29]. Raita et al. classified reason for visit according to Reason for Visit Classification of Diseases [26]. Sterling et al. used all individual words, paragraph vectors and topic extraction [30]. KATE extracts medical terms from free text using developed C-NLP technology and exploits them in predictive modeling. C-NLP demonstrates accuracy 98.47%, F1 score 0.992, sensitivity 0.997 and precision 0.9877 in extracting medical terms from free text (Table 2 and Supplementary Table 1). Extracting medical terms using C-NLP provides a more accurate description of patient presentations and medical history than categorization or simple use of individual words.

Since nurse error rate in triage reaches 40% it is not possible to accurately validate performance of machine learning models based on nurse labels. In contrast to previous studies, KATE was validated on a gold set in which all ESI labels were independently verified by ESI trained clinicians, providing confidence in KATE performance results.

Given the importance of accurate triage acuity assignment regarding the patient's clinical trajectory, improvement in triage accuracy has the potential to translate into better allocation of resources, more appropriate patient flow, and most importantly, more rapid identification of patients needing immediate care.

Although the main aim of this study was to evaluate nurses, expert clinicians, and KATE performance in ESI triage, we also analyzed the distribution of disposition per ESI for gold set (Supplementary Table 12).

Among patients that were admitted to hospital 36.05% were assigned ESI 1 or 2 by nurses, 61.63% by expert clinicians and 60.47% by KATE. This indicates the promise of KATE to improve patient outcomes by accurately assigning ESI for high acuity patients. In addition to model improvements for pediatric patients, further prospective research will focus on the impact of KATE on patient and operational outcomes.

## 5. Implications for Emergency Clinical Practice

Triage accuracy is critical to the process of getting patients to resources in a timely manner to ensure safe patient care. Our findings that triage acuity scores assigned by nurses were often inaccurate suggests that multiple factors impede accuracy in triage. The use of KATE, a clinical decision support aid, may facilitate this process and improve the initial clinical decision regarding acuity.

## 6. Conclusions

KATE, an ML model, has been determined to provide a triage acuity assignment substantially more accurate than the triage nurses in this study sample. Specifically, KATE operates independently of contextual factors, potentially mitigating the effects of implicit bias. KATE acuity score is based on many pieces of information drawn from the patient's medical history, medication history, and documented risk factors, along with vital signs and physiologic or psychological complaints. KATE is unaffected by the external pressures that can lead to under triage and mitigates the racial and social biases that can negatively affect the accuracy of triage assignment.

Future research should focus on the evaluation and improvement of KATE performance for critically ill children, the impact of KATE providing feedback to triage nurses in real time, and KATE's impact on mortality and morbidity, ED throughput, resource optimization, and nursing outcomes including competence, satisfaction and retention.

## 7. Limitations

This was a retrospective study, thus the contextual aspects of the triage process were not available for consideration. Similarly, the demographics of emergency nurses performing triage at the time of data input were not available. Although formal triage education was available at both sites, not all emergency nurses had taken that educational opportunity. The accuracy of the triage acuity assignments is congruent with reported accuracy in the literature, however it is possible that another cohort of triage nurses might have performed differently. For each individual high-risk presentation, a larger random sample gold set would need to be created to fully analyze KATE's performance on each of these presentations. For the pediatric population, records for children less than one year of age were not utilized, and there were not enough records for critically ill children to adequately assess the model performance.

## References

1. Centers for Disease Control and Prevention (CDC). National Center for Health Statistics (US). Health, United States, 2012: With Special Feature on Emergency Care. 2013; Report No.: 2013-1232.

2. Centers for Disease Control and Prevention. National Hospital Ambulatory Medical Care Survey: 2016 Emergency Department Summary Tables. 2016

3. Gilboy N, Tanabe P, Travers D, Rosenau A. Emergency Severity Index (ESI): A Triage Tool for Emergency Department Care, Version 4. Implementation Handbook 2012. Edition. Rockville: Agency for Healthcare Research and Quality, United States Department of Health & Human Services.

4. McHugh M, Tanabe P, McClelland M, Khare RK.  More patients are triaged using the Emergency Severity

Index than any other triage acuity system in the United States. Acad Emerg Med 2012 Jan;19(1):106-9. doi: 10.1111/j.1553-2712.2011.01240.x.

5. Croskerry P. Clinical cognition and diagnostic error: applications of a dual process model of reasoning. Adv Health Sci Educ Theory Pract 2009 Sep 1;14(1):27-35.

6. Saposnik G, Redelmeier D, Ruff CC, Tobler PN. Cognitive biases associated with medical decisions: a systematic review. BMC Med Inform Decis Mak 2016 Dec;16(1):138.

7. Wolf LA. Acuity assignation: an ethnographic exploration of clinical decision making by emergency nurses at initial patient presentation. Adv Emerg Nurs J 2010; 32: 234-246.

8. Wolf LA. An integrated, ethically driven environmental model of clinical decision making in emergency settings. Int J Nurs Knowl 2011; 24: 49-53.

9. Wolf LA, Delao AM, Perhats C, Moon MD, Zavotsky KE. Triaging the emergency department, not the patient: United States emergency nurses' experience of the triage process. J Emerg Nurs 2018; 44: 258-266.

10. Fry M, Burr G. Current triage practice and influences affecting clinical decision-making in emergency departments in NSW, Australia. Accid Emerg Nurs 2001; 9(4):227-34.

11. Chung JY. An exploration of accident and emergency nurse experiences of triage decision making in Hong Kong. Accid Emerg Nurs 2005;13:206–213

12. Saban M, Zaretsky L, Patito H, Salama R, Darawsha A. Round-off decision-making: Why do triage nurses assign STEMI patients with an average priority?. Int emerg nurs 2019 Mar 1;43:34-9.

13. Arslanian-Engoren C. Gender and age bias in triage decisions. J Emer Nurs 2000 Apr 1;26(2):117-24.

14. López L, Wilper AP, Cervantes MC, Betancourt JR, Green AR. Racial and sex differences in emergency

department triage assessment and test ordering for chest pain, 1997–2006. Acad Emerg Me 2010; 17: 801-808.

15. Schrader CD, Lawrence LM. Racial disparity in emergency department triage. J Emerg Med 2013; 44: 511-518.

16. Puumala SE, Burgess KM, Kharbanda AB, Zook HG, Castille DM, Pickner WJ, et al. The role of bias by emergency department providers in care for American Indian children. Med Care 2016; 54: 562-569.

17. Vigil JM, Alcock J, Coulombe P, McPherson L, Parshall M, Murata A, et al. Ethnic disparities in emergency severity index scores among US Veteran's affairs emergency department patients. PloS One 2015; 10.

18. Zook HG, Kharbanda AB, Flood A, Harmon B, Puumala SE, Payne NR. Racial differences in pediatric emergency department triage scores. J Emerg Med 2016; 50: 720-727.

19. Yurkova I, Wolf L. Under-triage as a significant factor affecting transfer time between the emergency department and the intensive care unit. J Emerg Nurs 2011; 37: 491–496.

20. Mistry B, Stewart De Ramirez S, Kelen G, Schmitz PSK, Balhara KS, Levin S, et al. Accuracy and Reliability of Emergency Department Triage Using the Emergency Severity Index: An International Multicenter Assessment. Ann Emerg Med 2018; 5: 581–587

21. Jordi K, Grossmann F, Gaddis GM, Cignacco E, Denhaerynck K, Schwendimann R, et al. Nurses' Accuracy and Self-Perceived Ability Using the Emergency Severity Index Triage Tool: A Cross-Sectional Study in Four Swiss Hospitals. Scand J Trauma Resusc Emerg Med 2015; 62.

22. Silva JAD, Emi AS, Leão ER, Lopes MCBT, Okuno MFP, Batista REA. Emergency Severity Index: Accuracy in Risk Classification. Einstein (Sao Paulo) 2017; 421–427.

23. Kwon JM, Lee Y, Lee Y, Lee S, Park H, Park J. Validation of deep-learning-based triage and acuity score using a large national dataset. PLoS One 2018; 13.

24. Rajkomar A, Oren E, Chen K, Dai AM, Hajaj N, Hardt M, et al. Scalable and accurate deep learning with electronic health records. NPJ Digit Med 2018; 18.

25. Tang F, Xiao C, Wang F, Zhou J. Predictive modeling in urgent care: a comparative study of machine learning approaches. JAMIA Open 2018; 1-12.

26. Raita Y, Goto T, Faridi MK, Brown DFM, Camargo CA Jr, Hasegawa K. Emergency department triage prediction of clinical outcomes using machine learning models. Crit Care 2019; 23: 64.

27. Goto T, Camargo CA Jr, Faridi MK, Freishtat RJ, Hasegawa K. Machine learning–based prediction of clinical outcomes for children during emergency department triage. JAMA Open Netw 2019; 2.

28. Zhang X, Kim J, Patzer RE, Pitts SR, Patzer A, Schrager JD. Prediction of Emergency Department Hospital Admission Based on Natural Language Processing and Neural Networks. Methods Inf Med 2017; 56: 377-389.

29. Hong WS, Haimovich AD, Taylor RA. Predicting hospital admission at emergency department triage using machine learning. PloS One 2018; 13.

30. Sterling NW, Patzer RE, Di M, Schrager JD. Prediction of emergency department patient disposition based on natural language processing of triage notes. Int J Med Informatics 2019; 129 184-188.

31. Dugas AF, Kirsch TD, Toerper M, Korley F, Yenokyan G, France D, et al. An electronic emergency triage system to improve patient distribution by critical outcomes. J Emerg Med 2016; 50: 910–918.

32. Gul M, Celic E. An exhaustive review and analysis on applications of statistical forecasting in hospital emergency departments. Health Systems 2018: 1-22.

33. Levin S, Toerper M, Hamrock E, Hinson JS, Barnes S, Gardner H, et al. Machine-Learning-Based Electronic Triage More Accurately Differentiates Patients With Respect to Clinical Outcomes Compared With the Emergency Severity Index. Ann Emerg Med 2017; 71: 565-574.

34. Bodenreider O. The Unified Medical Language System (UMLS): integrating biomedical terminology. Nucleic Acids Res 2004; 32: D267-70.

35. Chen T, Guestrin C. XGBoost: A Scalable Tree Boosting System. 2016; arXiv:1603.02754v3

36. Pishbin E, Ebrahimi M, Mirhaghi, A. Do physicians and nurses agree on triage levels in the emergency department? A meta-analysis. Notfall Rettungsmed 22, 379–385 (2019). https://doi.org/10.1007/s10049-019-0580-6

37. Hinson JS, Martinez DA, Schmitz PS, Toerper M, Radu D, Scheulen J, et al. Accuracy of emergency department triage using the Emergency Severity Index and independent predictors of under-triage and over-triage in Brazil: a retrospective cohort analysis. Int J Emerg Med 2018; 11: 3.

38. Andersson AK, Omberg M, Svedlund M. Triage in the emergency department – a qualitative study of the factors which nurses consider when making decisions. Nurs Crit Care 2006; 11: 136–145.

39. Vyas DA, Eisenstein LG, Jones DS. Hidden in plain sight—reconsidering the use of race correction in clinical algorithms. N Engl J Med 2020; 383:874-882 DOI: 10.1056/NEJMms2004740

40. Nugus P, Holdgate A, Fry M, Forero R, McCarthy S, Braithwaite J. Work pressure and patient flow management in the emergency department: findings from an ethnographic study. Acad Emerg Med 2011 Oct;18(10):1045-52.

41. Brosinski CM, Riddell AJ, Valdez S. Improving triage accuracy: a staff development approach. Clin Nurse Spec 2017; 31: 145-148.

**Supplementary Table 1.** Performance and 95% confidence intervals (in parentheses) of C-NLP for 800 randomly sampled medical records by clinical term type.

| Clinical term type | Number of clinical terms | Accuracy | F1 score | Sensitivity | Precision |
|---|---|---|---|---|---|
| Orientation | 392 | 1 | 1 | 1 | 1 |
| Primary pain onset | 52 | 1 | 1 | 1 | 1 |
| Reason for visit | 4391 | 0.9909 (0.9879 - 0.9936) | 0.9954 (0.9939 - 0.9968) | 0.9991 (0.9982 - 0.9998) | 0.9918 (0.9891 - 0.9943) |
| Previous illness | 687 | 0.9869 (0.9767 - 0.9942) | 0.9934 (0.9882 - 0.9971) | 0.9883 (0.9782 - 0.9956) | 0.9985 (0.9942 - 1.0) |
| Surgeries | 1063 | 0.9454 (0.9304 - 0.9586) | 0.972 (0.9639 - 0.9789) | 0.996 (0.992 - 0.999) | 0.949 (0.9346 - 0.9613) |
| Primary pain quality | 39 | 1 | 1 | 1 | 1 |
| Primary pain location | 436 | 0.9954 (0.9885 - 1.0) | 0.9977 (0.9942 - 1.0) | 0.9954 (0.9885 - 1.0) | 1 |
| Level of consciousness | 789 | 1 | 1 | 1 | 1 |
| Affect behavior | 33 | 1 | 1 | 1 | 1 |
| Primary pain location detail | 69 | 1 | 1 | 1 | 1 |
| Prior to arrival | 109 | 0.9817 (0.9541 - 1.0) | 0.9907 (0.9765 - 1.0) | 1.0 (1.0 - 1.0) | 0.9817 (0.9541 - 1.0) |
| Respiratory status | 26 | 1 | 1 | 1 | 1 |
| Triage treatment | 370 | 0.9838 (0.9703 - 0.9946) | 0.9918 (0.9849 - 0.9973) | 0.9838 (0.9703 - 0.9946) | 1 |
| Family history | 63 | 1 | 1 | 1 | 1 |
| Problems | 928 | 0.9698 (0.9569 - 0.9795) | 0.9847 (0.978 - 0.9897) | 0.9956 (0.9901 - 0.9989) | 0.974 (0.9623 - 0.9838) |
| Menstrual | 47 | 1 | 1 | 1 | 1 |
| Primary pain radiation location | 5 | 1 | 1 | 1 | 1 |
| Primary pain radiation location detail | 2 | 1 | 1 | 1 | 1 |
| Primary pain aggravating factors | 1 | 1 | 1 | 1 | 1 |
| Primary pain associated | 1 | 1 | 1 | 1 | 1 |
| symptoms | | | | | |
| Medical devices | 3 | 1 | 1 | 1 | 1 |

**Supplementary Table 2.** Confusion matrix of Nurse ESI against verified ESI for Study sites A and B

| ESI Label | Verified ESI 1 | Verified ESI 2 | Verified ESI 3 | Verified ESI 4 | Verified ESI 5 | Total count |
|---|---|---|---|---|---|---|
| Nurse ESI 1 | 117 | 68 | 16 | 9 | 2 | 212 |
| Nurse ESI 2 | 553 | 1484 | 484 | 73 | 30 | 2624 |
| Nurse ESI 3 | 122 | 4991 | 4124 | 700 | 573 | 10510 |
| Nurse ESI 4 | 8 | 412 | 715 | 1856 | 2705 | 5696 |
| Nurse ESI 5 | 0 | 30 | 28 | 163 | 389 | 610 |
| Total count | 800 | 6985 | 5367 | 2801 | 3699 | 19652 |

**Supplementary Table 3.** Comparison of accuracy scores with 95% confidence intervals (in parentheses) for KATE with and without verified ESI labels

| Group | Number of triage records | KATE accuracy without verified ESI labels (95% CI) | KATE accuracy with verified ESI labels (95% CI) |
|---|---|---|---|
| All records | 729 | 0.711 (0.676 - 0.739) | 0.759 (0.724 - 0.787) |
| Site A | 368 | 0.755 (0.709 - 0.796) | 0.799 (0.758 - 0.834) |
| Site B | 361 | 0.665 (0.615 - 0.707) | 0.717 (0.665 - 0.759) |
| All ESI 1 | 5 | 0.2 (0.0 - 0.6) | 0.6 (0.2 - 1.0) |
| All ESI 2 | 145 | 0.683 (0.6 - 0.752) | 0.8 (0.724 - 0.869) |
| All ESI 3 | 277 | 0.823 (0.773 - 0.866) | 0.827 (0.776 - 0.87) |
| All ESI 4 | 210 | 0.757 (0.7 - 0.809) | 0.757 (0.695 - 0.814) |
| All ESI 5 | 92 | 0.337 (0.239 - 0.424) | 0.5 (0.391 - 0.598) |

| Supplementary Table 4. Distribution of reasoning for record removal from the study gold set. Of the initial 800 records sample, 71 records were removed. | |
|---|---|
| Conflicting documentation | 7 (9.86%) |
| Clinical team could not reach consensus | 19 (26.76%) |
| Impossible vital signs | 2 (2.82%) |
| Insufficient information | 31 (43.66%) |
| Missing 4 or more vitals signs | 7 (9.86%) |
| No reason for visit | 5 (7.04%) |

| Supplementary Table 5. ESI acuity distribution for each group in the gold set (n=729) | | | | | | |
|---|---|---|---|---|---|---|
| **ESI Label** | **Nurse** | **Gold** | **Clinician 1** | **Clinician 2** | **Clinician 3** | **KATE** |
| ESI 1 | 1 (0.14%) | 5 (0.69%) | 3 (0.41%) | 3 (0.41%) | 2 (0.27%) | 3 (0.41%) |
| ESI 2 | 88 (12.07%) | 145 (19.89%) | 129 (17.70%) | 134 (18.38%) | 134 (18.38%) | 144 (19.75%) |
| ESI 3 | 363 (49.79%) | 277 (38.00%) | 293 (40.19%) | 295 (40.47%) | 295 (40.47%) | 308 (42.25%) |
| ESI 4 | 241 (33.06%) | 210 (28.81%) | 238 (32.65%) | 242 (33.20%) | 214 (29.36%) | 213 (29.22%) |
| ESI 5 | 36 (4.94%) | 92 (12.62%) | 66 (9.05%) | 55 (7.54%) | 78 (10.70%) | 61 (8.37%) |
| No ESI label | - | - | - | - | 6 (0.82%) | - |
| Group descriptions: Nurse - ESI acuity assigned by nurse in study dataset  Gold - Consensus ESI acuity by clinicians 1 ,2, 3  Clinician 1, 2, 3 - Each individual study clinician initial ESI acuity, before consensus agreement  KATE - ML model acuity prediction | | | | | | |

| Supplementary Table 6. Micro-average AUC with 95% confidence intervals (in parentheses) for all age groups (pediatric & adult patients) in the gold set | | | | | | |
|---|---|---|---|---|---|---|
| **Group** | **Number of triage records** | **KATE AUC (95% CI)** | **Nurse AUC (95% CI)** | **Clinician AUC (95% CI)** | **Clinician 2 AUC (95% CI)** | **Clinician 3 AUC (95% CI)** |
| All records | 729 | 0.849 (0.828 - 0.867) | 0.749 (0.725 - 0.769) | 0.86 (0.84 - 0.877) | 0.855 (0.835 - 0.871) | 0.822 (0.802 - 0.841) |
| Site A | 368 | 0.866 (0.839 - 0.889) | 0.764 (0.73 - 0.797) | 0.83 (0.797 - 0.857) | 0.817 (0.784 - 0.846) | 0.812 (0.778 - 0.842) |
| Site B | 361 | 0.823 (0.79 - 0.849) | 0.718 (0.683 - 0.746) | 0.881 (0.856 - 0.905) | 0.882 (0.855 - 0.906) | 0.82 (0.788 - 0.846) |

**Supplementary Table 7.** Macro-average F1-score with 95% confidence intervals (in parentheses) for all age groups (pediatric & adult patients) in the gold set

| Group | Number of triage records | KATE F1-score (95% CI) | Nurse F1-score (95% CI) | Clinician 1 F1-score (95% CI) | Clinician 2 F1-score (95% CI) | Clinician 3 F1-score (95% CI) |
|---|---|---|---|---|---|---|
| All records | 729 | 0.738 (0.585 - 0.802) | 0.428 (0.394 - 0.459) | 0.757 (0.606 - 0.817) | 0.74 (0.588 - 0.797) | 0.665 (0.538 - 0.746) |
| Site A | 368 | 0.766 (0.714 - 0.812) | 0.56 (0.499 - 0.614) | 0.664 (0.602 - 0.721) | 0.595 (0.542 - 0.64) | 0.631 (0.57 - 0.683) |
| Site B | 361 | 0.705 (0.551 - 0.77) | 0.397 (0.345 - 0.439) | 0.799 (0.648 - 0.863) | 0.796 (0.647 - 0.861) | 0.68 (0.543 - 0.764) |

**Supplementary Table 8.** Macro-average sensitivity with 95% confidence intervals (in parentheses) for all age groups (pediatric & adult patients) in the gold set

| Group | Number of triage records | KATE sensitivity (95% CI) | Nurse sensitivity (95% CI) | Clinician 1 sensitivity (95% CI) | Clinician 2 sensitivity (95% CI) | Clinician 3 sensitivity (95% CI) |
|---|---|---|---|---|---|---|
| All records | 729 | 0.695 (0.573 - 0.786) | 0.417 (0.389 - 0.445) | 0.714 (0.593 - 0.801) | 0.697 (0.577 - 0.788) | 0.626 (0.532 - 0.735) |
| Site A | 368 | 0.764 (0.707 - 0.812) | 0.547 (0.488 - 0.601) | 0.644 (0.589 - 0.693) | 0.601 (0.56 - 0.641) | 0.625 (0.573 - 0.67) |
| Site B | 361 | 0.657 (0.527 - 0.752) | 0.392 (0.352 - 0.429) | 0.771 (0.654 - 0.862) | 0.768 (0.645 - 0.858) | 0.655 (0.555 - 0.758) |

**Supplementary Table 9.** Macro-precision sensitivity with 95% confidence intervals (in parentheses) for all age groups (pediatric & adult patients) in the gold set

| Group | Number of triage records | KATE precision (95% CI) | Nurse precision (95% CI) | Clinician 1 precision (95% CI) | Clinician 2 precision (95% CI) | Clinician 3 precision (95% CI) |
|---|---|---|---|---|---|---|
| All records | 729 | 0.809 (0.606 - 0.836) | 0.488 (0.446 - 0.528) | 0.83 (0.627 - 0.854) | 0.822 (0.62 - 0.85) | 0.757 (0.544 - 0.785) |
| Site A | 368 | 0.769 (0.712 - 0.818) | 0.601 (0.527 - 0.675) | 0.733 (0.65 - 0.799) | 0.687 (0.543 - 0.825) | 0.672 (0.594 - 0.748) |
| Site B | 361 | 0.805 (0.598 - 0.844) | 0.507 (0.446 - 0.562) | 0.85 (0.657 - 0.883) | 0.846 (0.639 - 0.877) | 0.761 (0.543 - 0.796) |

**Supplementary Table 10.** ESI acuity assignment accuracy with 95% confidence intervals (in parentheses) for adult patients in the gold set

| Group | Number of triage records | KATE model accuracy (95% CI) | Nurse triage accuracy (95% CI) | Clinician 1 accuracy (95% CI) | Clinician 2 accuracy (95% CI) | Clinician 3 accuracy (95% CI) |
|---|---|---|---|---|---|---|
| All records | 562 | 0.785 (0.749 - 0.817) | 0.616 (0.573 - 0.653) | 0.81 (0.774 - 0.838) | 0.783 (0.747 - 0.815) | 0.748 (0.708 - 0.783) |
| Site A | 272 | 0.82 (0.772 - 0.86) | 0.647 (0.588 - 0.699) | 0.783 (0.728 - 0.827) | 0.754 (0.702 - 0.805) | 0.79 (0.739 - 0.835) |
| Site B | 290 | 0.752 (0.7 - 0.797) | 0.586 (0.528 - 0.638) | 0.834 (0.79 - 0.869) | 0.81 (0.762 - 0.852) | 0.707 (0.652 - 0.756) |
| All ESI 1 | 5 | 0.6 (0.2 - 1.0) | 0.0 (0.0 - 0.0) | 0.6 (0.2 - 1.0) | 0.6 (0.2 - 1.0) | 0.4 (0.0 - 0.8) |
| All ESI 2 | 131 | 0.855 (0.794 - 0.916) | 0.427 (0.343 - 0.511) | 0.786 (0.71 - 0.855) | 0.786 (0.71 - 0.855) | 0.746 (0.669 - 0.815) |
| All ESI 3 | 251 | 0.841 (0.793 - 0.884) | 0.797 (0.741 - 0.845) | 0.837 (0.789 - 0.876) | 0.849 (0.805 - 0.892) | 0.803 (0.747 - 0.851) |
| All ESI 4 | 129 | 0.721 (0.643 - 0.798) | 0.62 (0.527 - 0.698) | 0.837 (0.767 - 0.899) | 0.791 (0.713 - 0.853) | 0.721 (0.643 - 0.791) |
| All ESI 5 | 46 | 0.478 (0.326 - 0.609) | 0.217 (0.109 - 0.348) | 0.674 (0.543 - 0.804) | 0.413 (0.261 - 0.565) | 0.565 (0.413 - 0.696) |

**Supplementary Table 11.** ESI acuity assignment accuracy with 95% confidence intervals (in parentheses) for pediatric patients in the gold set

| Group | Number of triage records | KATE accuracy (95% CI) | Nurse triage accuracy (95% CI) | Clinician 1 accuracy (95% CI) | Clinician 2 accuracy (95% CI) | Clinician 3 accuracy (95% CI) |
|---|---|---|---|---|---|---|
| All records | 167 | 0.671 (0.599 - 0.743) | 0.539 (0.455 - 0.617) | 0.665 (0.587 - 0.731) | 0.719 (0.647 - 0.784) | 0.604 (0.524 - 0.677) |
| Site A | 96 | 0.74 (0.646 - 0.823) | 0.646 (0.541 - 0.729) | 0.635 (0.531 - 0.729) | 0.646 (0.552 - 0.729) | 0.511 (0.404 - 0.606) |
| Site B | 71 | 0.577 (0.465 - 0.676) | 0.394 (0.268 - 0.507) | 0.704 (0.592 - 0.803) | 0.817 (0.718 - 0.901) | 0.729 (0.614 - 0.814) |
| All ESI 1 | 0 | - | - | - | - | - |
| All ESI 2 | 14 | 0.286 (0.071 - 0.5) | 0.286 (0.071 - 0.5) | 0.571 (0.356 - 0.786) | 0.714 (0.5 - 0.929) | 0.857 (0.643 - 1.0) |
| All ESI 3 | 26 | 0.692 (0.5 - 0.846) | 0.5 (0.308 - 0.692) | 0.769 (0.577 - 0.923) | 0.808 (0.654 - 0.923) | 0.769 (0.577 - 0.885) |
| All ESI 4 | 81 | 0.815 (0.716 - 0.889) | 0.765 (0.667 - 0.852) | 0.765 (0.667 - 0.852) | 0.802 (0.716 - 0.876) | 0.605 (0.506 - 0.704) |
| All ESI 5 | 46 | 0.522 (0.37 - 0.652) | 0.239 (0.109 - 0.348) | 0.457 (0.304 - 0.587) | 0.522 (0.37 - 0.652) | 0.419 (0.256 - 0.558) |

| **Supplementary Table 12.** Distribution of disposition for gold set by assigned ESI by nurses, expert clinicians and KATE | | | | | |
|---|---|---|---|---|---|
| **Nurses** | | | | | |
| Disposition | ESI 1 | ESI 2 | ESI 3 | ESI 4 | ESI 5 |
| Discharge | 1 (0.16%) | 50 (8.16%) | 293 (47.80%) | 233 (38.01%) | 36 (5.87%) |
| Admit | 0 (0%) | 31 (36.05%) | 52 (60.47%) | 3 (3.49%) | 0 (0%) |
| **Expert clinicians** | | | | | |
| Discharge | 1 (0.16%) | 88 (14.36%) | 233 (38.01%) | 202 (32.95%) | 89 (14.52%) |
| Admit | 4 (4.65%) | 49 (56.98%) | 28 (32.56%) | 3 (3.49%) | 2 (2.33%) |
| **KATE** | | | | | |
| Discharge | 0 (0%) | 82 (13.38%) | 267 (43.56%) | 205 (33.44%) | 59 (9.62%) |
| Admit | 3 (3.49%) | 49 (56.98%) | 32 (37.21%) | 2 (2.33%) | 0 (0%) |

## COI Declarations

- Ethics approval and consent to participate
  - This study was approved and exempted by the Western Institutional Review Board (WIRB, OHRP/FDA Parent Organization number: IORG0000432, IRB registration number: IRB00000533)
- Consent for publication
  - Not Applicable
- Availability of data and material
  - The datasets generated during and/or analyzed during the current study are not publicly available due to the fact they contain person health information from the medical records of the study.
- Competing interests
  - Work Under Consideration for Publication:
    - This work was conducted and independently funded by Mednition Inc. a private corporation.
    - All Authors that are with the Mednition organization are employees. All non Mednition authors are contract employees of Mednition with the exception of Robert Dunne, MD.
  - Relevant financial activities outside the submitted work:
    - None
  - Intellectual Property:
    - Mednition has submitted a provisional patent for the work related to the study
  - Relationships not covered:
    - None
- Funding
  - This work was conducted and independently funded by Mednition Inc. a private corporation.
- Authors' contributions

- o All authors read and approved the final manuscript
- o OI was the primary author of the papers machine learning and results content.
- o LW was the primary author of the clinical sections and content relating to triage
- o DB, KM, and EL reviewed and supported LW clinical content
- o KM, YA, and MM reviewed and analyzed data in support of the primary author (OI)
- o SL, RD, KK provided senior clinical review of the paper
- o CR was the study designer and lead researcher for the study
- Acknowledgements
  - o Not Applicable